\documentclass[useAMS,usenatbib]{mn2e}
\usepackage{graphics,epsfig,amsmath,amssymb,amstext,txfonts,array,color}
\usepackage{graphicx}
\usepackage[T1]{fontenc}

\usepackage{color}

\newcommand\degree{^{\circ}}

\newcommand\simlt{\lower.5ex\hbox{$\; \buildrel < \over \sim \;$}}
\newcommand\simgt{\lower.5ex\hbox{$\; \buildrel > \over \sim \;$}}

\title[The collimation of magnetic jets by disk winds]
{The collimation of magnetic jets by disk winds}

\author[Noemie Globus, Amir Levinson]
  {N.~Globus,$^1$\thanks{E-mail: noemie.globus@mail.huji.ac.il}
  A.~Levinson,$^2$  \\
  $^1$Racah Institute of Physics, The Hebrew University of Jerusalem, 91904 Jerusalem, Israel\\
  $^2$Raymond and Beverly Sackler School of Physics \& Astronomy, Tel Aviv University, Tel Aviv 69978, Israel}
\date{Released \today}

\pagerange{\pageref{firstpage}--\pageref{lastpage}} \pubyear{2002}

\def\LaTeX{L\kern-.36em\raise.3ex\hbox{a}\kern-.15em
    T\kern-.1667em\lower.7ex\hbox{E}\kern-.125emX}

\begin{document}
\label{firstpage}
\maketitle

\begin{abstract}
The collimation of a Poynting-flux dominated jet by a wind emanating from the surface of an accretion flow 
is computed using a semi-analytic model.  The injection of the disk wind is treated as a boundary condition
in the equatorial plane, and its evolution is followed by invoking a prescribed geometry of streamlines.   
Solutions are obtained for a wide range of disk wind parameters.   It is found that  jet collimation generally occurs 
when the total wind power exceeds about ten percents of the jet power.   For moderate wind powers we find gradual 
collimation.  For strong winds we find  rapid collimation followed by  focusing of the jet, after which it remains 
narrow  over many Alfv\'en crossing times before becoming conical.
We estimate that  in the later case the jet's magnetic field may be dissipated by 
the current-driven kink instability over a distance of a few hundreds gravitational radii.   We apply the model to M87
and show that the observed parabolic shape of the radio jet within the Bondi radius 
can be reproduced provided that the wind injection zone extends to several hundreds gravitational radii, and that
its total power is about one third of the jet power.    The radio spectrum can be produced by synchrotron radiation 
of relativistically hot, thermal electrons in the sheath flow surrounding the inner jet. 
\end{abstract} 

\begin{figure*}
\centerline{\includegraphics[scale=0.2]{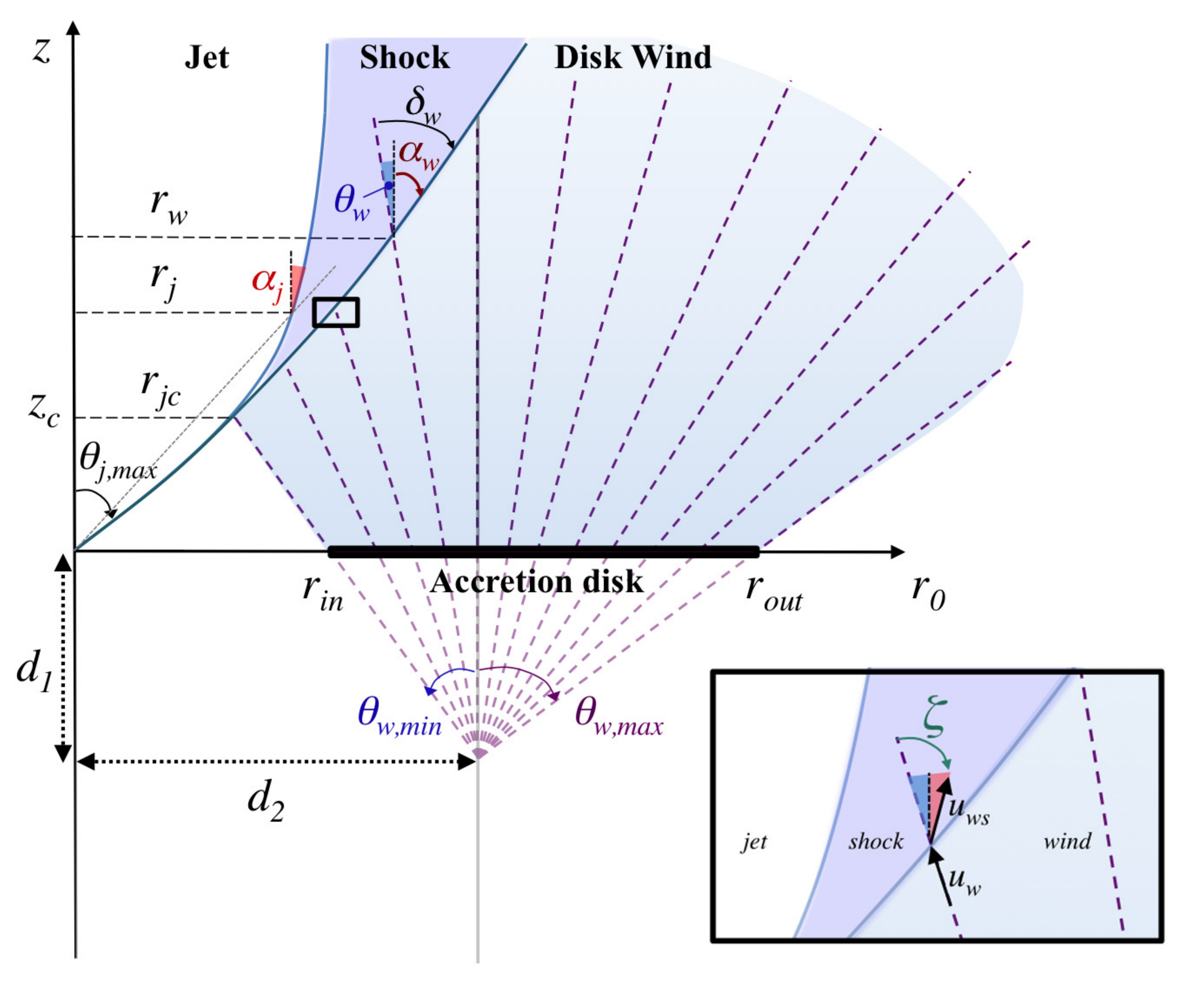}}
\caption{\label{f1} Sketch of the geometry of the colliding outflows model.  The jet and wind quantities are designated as $j$ and $w$ respectively.}
\end{figure*}

\section{Introduction}

Relativistic jets are ubiquitous in accreting black hole systems.  They provide a means by which the free energy of the 
central engine is transported to large distances, and ultimately converted to the observed radiation.    Although their composition
is not well constrained, it is commonly believed that they are dominated by magnetic fields, at least on sufficiently small scales.
Strong collimation appears to be a generic feature of relativistic jets, and there are various indications that collimation is accomplished relatively close to the black hole.   Collimation is intimately related
to the dynamics of Poynting dominated flows (e.g., Begelman \& Li 1994; Vlahakis 2004; Komissarov et al. 2009; Lyubarsky 2009) 
and conceivably to the dissipation of magnetic energy (e.g., Levinson \& Begelman 2013; Bromberg \& Tchekhovskoy 2016). Therefore, resolving 
the collimation mechanism is essential for the understanding of relativistic jets, their emission and their interaction with their environments. 

It has been shown that magnetic self-collimation is a fundamental property of rotating magnetized jets (Heyvaerts \& Norman 1989), because they gradually build-up a large scale toroidal magnetic field, with associated hoop stresses that act to collimate the flow. However, it is a slow process in non-relativistic flows (Eichler 1993), and even less effective in relativistic plasma flows, even at a very fast rotation (Tomimatsu 1994; Begelman \& Li 1994; 
Beskin et al. 1998; Okamoto 1999), and so cannot account for the inferred collimation scales. 
Confinement by the pressure and inertia of an external medium has emerged as a promising alternative.    The environments in which astrophysical jets propagate, e.g., accretion disk winds and  ambient gas clouds in the case of AGNs and micro-quasars, and stellar envelopes in the case of long gamma-ray bursts (GRBs), are ideal for this purpose (e.g., Eichler 1983; Begelman 1995; Levinson \& Eichler 2000; Bromberg \& Levinson 2007, 2009; Nalewajko \& Sikora 2009; Kohler et al. 2012).  

A jet propagating in a dense ambient medium is  surrounded by a hot cocoon that forms due to side flows of shocked matter from the jet's head, or a nose cone if magnetic pinching is important (Komissarov 1999).  Analytic models and numerical simulations of purely hydrodynamic jets
(Begelman \& Cioffi 1989; Marti et al. 1997; Aloy et al. 1999; Hughes et al. 2002; Matzner 2003; Zhang et al. 2003; 
Lazzati \& Begelman 2005;  Morsony et al. 2007; Mizuta \& Aloy 2009; Lazzati et al. 2009; Bromberg et al. 2011) as well as 
magnetic jets (Lind et al. 1989; Van Putten 1996; Komissarov 1999; Levinson \& Begelman 2013; Bromberg et al. 2014; Bromberg \& Tchekhovskoy 2016)  indicate that under astrophysical conditions anticipated in GRBs, active galactic nuclei (AGNs) and microquasars, the surrounding cocoon can collimate and confine the jet
during certain phases of its evolution.  The details depend primarily on the density profile of the ambient medium.  In GRBs, the over-pressured 
cocoon can effectively confine the jet prior to its breakout from the stellar envelope. In case of Poynting-flux dominated jets, the strong 
confinement can lead to rapid dissipation of the magnetic field at the collimation nozzle (Bromberg \& Tchekhovskoy 2016). 
After jet breakout from the stellar envelope the cocoon fades away, roughly over the sound crossing time, and eventually disappears.    Whether it can survive for long enough time to keep the outflow collimated as it reaches the photosphere, which is located typically at a distance of tens to hundreds stellar radii, is yet an open issue.   If not, an alternative mechanism is needed to produce the beaming inferred during the 
emission phase.   Disk winds, if sufficiently powerful, is a viable alternative (Levinson \& Eichler 2000, Bromberg \& Levinson 2007).     
Powerful AGN jets can be confined by their cocoons on kpc scales and beyond (Bromberg et al. 2011).
This cannot explain the high degree of collimation observed on sub-parsec scales in many sources, that imply the presence of a 
confining medium in the vicinity of the black hole.   In M87, for instance, the radio
jet appears to have a parabolic profile over the range of  radii $50\, r_s$ to $10^5\, r_s$, above which it becomes conical (Asada \& Nakamura 2012).  
This requires the presence of ambient gas with a pressure profile $p_{ext}\propto r^{-2}$ in the collimation zone (Lyubarsky 2009),   
the origin of which is unclear.  In section \ref{sec:M87} we show that such a profile can be  produced by an extended disk wind. 

Substantial winds from the surface of the accretion flow have been predicted for accretion rates well below and well 
above the Eddington
rate (Blandford \& Begelman 1999, 2004; Begelman 2012).   Recent numerical simulations  (Moscibrodzka \& Falcke 2013;  
Sadowski \& Narayan 2015, 2016) confirm  these predictions.   In case of a rapidly rotating black hole the simulations indicate 
a multi-flow  consisting of an inner, highly magnetized relativistic jet surrounded by a denser, subrelativistic flow.  However, 
given the limited size of the simulation box, the flow cannot be followed to large distances. 
Moreover, in most of the cases reported the different components are not well resolved. 
 In this paper we study the interaction of a disk wind with a Poynting-flux dominated jet launched by the putative black hole 
using a semi-analytic approach. 
Our main purpose is to compute the profile of the magnetic jet for a range of wind properties.  We find that quite generally,
jet collimation requires a fraction of total wind and jet powers of $L_w/L_j > 0.1$.  The collimation radius depends, 
in addition, on the horizontal extent and power distribution of the disk wind.  
For relatively powerful winds, $L_w\sim L_j$, the jet passes through a collimation nozzle before becoming conical.
We find that in this strong collimation 
regime the jet is kink unstable, and speculate that in such circumstances magnetic dissipation may occur on scales
of several hundreds Schwarzschild radii. 

The plan of the paper is as follows:   Sect. 2 outlines the colliding wind-jet model.  Sect. 3 describes the main results. 
In Sect. 4 we discuss the stability of magnetic jets confined by strong disk winds.  In Sect. 5 we apply the model to the jet in M87
and show that its collimation profile can be explained by an extended disk wind.  We also estimate the synchrotron 
luminosity from the sheath outflow surrounding the jet.   We conclude in Sect. 6. 

\section{The Model}

We construct a class of semi-analytic models for the collimation and confinement of a Poynting-flux dominated jet by
a sub-relativistic, hydrodynamic wind emanating from the upper layers of the accretion disk surrounding the black hole. 
The resultant multi-flow structure depends on whether the collision of the disk wind and the inner jet is subsonic or 
supersonic.   In the latter case it consists of the inner jet, the outer wind, and a shocked wind layer enclosed between the 
contact surface and an oblique shock across which the streamlines of the disk wind are deflected,  as shown schematically in 
Figure \ref{f1}.    In the case of a subsonic collision the inner jet  is also expected to be ensheathed by a slow, compressed flow,
established through a smooth deflection of the outer wind streamlines.   Our analysis is restricted to situations where the collision 
is initially supersonic, albeit becoming subsonic as the flow expands, 
as we were unable to construct a reasonable semi-analytic model of subsonic confinement in cases where the initial
opening angle of the inner jet is not small.  It seems that numerical simulations are required for a detailed investigation of this regime. 
Nonetheless, since the collimation and confinement of the inner jet depend only on the pressure profile at its boundary, we believe 
that our results provide a good qualitative description also in the regime of subsonic collision. 
In our treatment, the unshocked disk wind is given as an input, and we use a parametrised model in order to obtain the local conditions
just upstream of the oblique shock, as explained in detail below.    For the inner, Poynting-flux dominated jet we employ the model 
developed by  Lyubarsky  (2009, 2011).  Since the profile of the inner jet depends on the pressure distribution in
the shocked wind layer, the equation describing the evolution of the inner jet is coupled to the equations governing the structure of the
shocked wind layer.   This set of equations is subject to boundary conditions at the shock front, that depend on the local properties of
the flow just upstream of the shock.  For a given input disk wind model, this set of coupled equations can be solved to obtain 
the entire multi-flow structure. 
Below we present stationary, axisymmetric solutions of these equations for a range of disk wind models.

We use cylindrical coordinates ($r$,$\phi$,$z$), with the $z$ axis aligned with the jet symmetry axis.  The cylindrical radius 
measured at the disk midplane ($z=0$) is denoted by $r_0$. 
The wind injection zone in the disk midplane extends from $r_0=r_{\rm in}$ to $r_0=r_{\rm out}$.  
In what follows, subscripts $j$, $w$ and $ws$ refer to quantities measured in the inner jet, 
the unshocked disk wind and the shocked wind layer, respectively. The various zones are indicated in Figure \ref{f1}.
We denote the density, pressure, and velocity field of the outer wind by $\rho_w(z,r)$, $p_w(z,r)$, and $\vec{u}_w(z,r)=
u_{w,z}(z,r)\hat{z}+u_{w,r}(z,r)\hat{r}$, respectively, and assume adiabatic flow, $\vec{u}_w\cdot \nabla(p_w/\rho_w^{\hat{\gamma}})=0$.  
The shocked wind layer is contained between the jet boundary, $r_j(z)$, and the shock surface, $r_w(z)$, which are unknown 
a priori.  Its density $\rho_{ws}(z,r)$, pressure $p_{ws}(z,r)$, and velocity $\vec{u}_{ws}(z,r)=u_{ws,z}(z,r)\hat{z}+u_{ws,r}(z,r)\hat{r}$, 
satisfy the continuity, energy and momentum equations, subject to the following boundary conditions:  
At the contact, $r_j(z)$, the velocity must be tangent to the jet boundary; that is, 
\begin{equation}
\frac{u_{ws,r}(z,r_j)}{u_{ws,z}(z,r_j)}=\frac{dr_j}{dz}\equiv \tan\alpha_j.
 \label{BC_j}
\end{equation}
At the shock front, $r_w(z)$, the shocked fluid quantities must satisfy the local jump conditions:
\begin{eqnarray}
\frac{\rho_{w}}{\rho_{ws}}&=&\frac{u_{ws||}}{u_{w||}}=\frac{2}{{\rm M}_w^2\sin^2\delta_w(\hat{\gamma}+1)}+\frac{\hat{\gamma}-1}{\hat{\gamma}+1}\,, \label{comp_r}\\
\frac{p_{ws}}{p_w}&=&\frac{2\hat{\gamma}}{\hat{\gamma}+1}{\rm M}_w^2\sin^2\delta_w-\frac{\hat{\gamma}-1}{\hat{\gamma}+1}\,, \label{p_ws}\\
u_{ws\perp}&=&u_{w\perp},
\end{eqnarray}
where $u_{w\perp}=u_{w}\cos\delta_w$ denotes the component of the unshocked wind velocity  perpendicular to the 
shock normal (i.e., tangent to the shock 
surface),  $u_{w||}=u_{w}\sin\delta_w$ the component along to the shock normal, $\delta_w$ the impact angle, and ${\rm M}_w$ the Mach
number of the unshocked wind. The magnitude of the fluid velocity just downstream of the shock can be
expressed in terms of the upstream wind velocity $u_w$, the impact angle $\delta_w$ and the compression ratio $r=\rho_{ws}/\rho_w$ as
\begin{eqnarray}
u_{ws}=\sqrt{u_{ws||}^2+u_{ws\perp}^2}=u_w\sqrt{\cos^2\delta_w+r^{-2}\sin^2\delta_w}. \label{u_ws}
\end{eqnarray}
>From Figure \ref{f1} it is seen that the slope of the shock surface satisfies
\begin{equation}
\frac{dr_w}{dz}=\tan(\theta_w+\delta_w),\label{dr_w/dz}
\end{equation}
where $\theta_w$ is the angle between the velocity of the unshoked wind and the jet axis, that is,
$\cos\theta_w=\hat{u}_{w}(z,r_w)\cdot\hat{z}$.    
For the wind geometry adopted in section \ref{outer_wind} below, it is given explicitly in the text above Equation (\ref{r0}).

A useful quantity, to be used later, is the deflection angle $\zeta$ of a fluid element crossing the shock front,
given explicitly by $\cos\zeta=\hat{u}_{ws}(z,r_w)\cdot\hat{u}_w(z,r_w)$ (see the inset in Figure \ref{f1}).  The
jump conditions yield the relation
\begin{equation}
\tan \zeta = 2 \cot \delta_w \frac{{\rm M}_w^2\sin^2\delta_w-1}{2+  {\rm M}_w^2 (\hat{\gamma}+\cos2\delta_w)} \,.
 \label{deflect}
\end{equation}
For a given choice of the deflection angle $\zeta$, Equation (\ref{deflect}) admits three roots for the impact angle $\delta_w$,  two
of which are physical (see appendix \ref{appA}).   One solution corresponds to a strong shock and one to a weak shock. 
The appropriate choice in our case is dictated by the asymptotic behaviour of the flows.    From Equation (\ref{p_ws}) it is evident that
an oblique shock will form locally  provided ${\rm M}_w^2\sin^2\delta_w>1$.   From Figure \ref{f1} it is seen that the angle between the streamlines 
of the disk wind and the inner jet surface decreases with increasing  $z$.  This implies that the shock weakens as the flow propagates outwards, and ultimately disappears once ${\rm M}_w^2\sin^2\delta_w$ approaches unity.   Thus, the appropriate choice is the weak shock solution.  We verified that the choice of a strong shock solution yields an unphysical behaviour.

\subsection{The inner jet}

We suppose that the inner jet is injected into a cone of opening angle $\theta_{j,max}$ (Figure \ref{f1}), with
a total power $L_j$. Initially, it undergoes free expansion, whereby $\tan\theta_{j,max}=r_j/z$, until the ram pressure 
of the wind colliding with it becomes comparable to the jet pressure, at which point the jet becomes confined by the 
pressure of the shocked disk wind layer.   From that point on the entire multi-flow structure depends on the response of the 
inner jet to the pressure of the shocked plasma, and an additional equation governing the evolution of the inner jet is needed.

Asymptotic equations describing  a strongly magnetized jet have been derived by Komissarov et al. (2009) and Lyubarsky (2009, 2011),
and applied to jets confined by external medium.     Lyubarsky obtained analytic solutions of the asymptotic transfield equation in the case of a rigidly rotating, steady jet confined by an external pressure having a power law profile $p_{ext}(z)\propto z^{-\kappa}$.  He examined the behavior of the solutions in the regimes $\kappa<2$ and $\kappa>2$, and showed that when $\kappa<2$  the opening angle of the jet $\theta_j$ decreases continuously such that the jet interior remains in causal contact ($\gamma_j\theta_j\simlt1$) everywhere.  As a consequence,  the jet is accelerated and collimated until it roughly reaches equipartition, where $\sigma\sim1$.  The jet's streamlines have a parabolic shape, $r\propto z^{\kappa/4}$,  with spatial oscillations superimposed on it if the jet is initially out of equilibrium.
 For $\kappa>2$ the jet becomes asymptotically radial, with the final opening angle $\theta_{j\infty}$ depending solely on the pressure profile.
 Our results, though more involved, are consistent with those trends. 

In the case considered here, the jet is confined by the pressure of the shocked wind $p_{ws}(z)$, which is unknown a priori.  Therefore,
the solutions obtained by Lyubarsky cannot be used directly.   Instead, we employ Equation (53) from Lyubarsky (2009) for the jet radius.
In our notation it reads:
\begin{equation}
\frac{d^2 r_j(z)}{dz^2}=\frac{r_L^2}{[r_j(z)]^3}-\frac{3\pi c\, p_{ws}(z) r_j(z)}{2 L_{j}}\,,
\label{asymptotic_jet}
\end{equation}
where $r_L$ is the radius of the light cylinder, and the jet power is related to the magnetic field at the light 
cylinder, $B_0$, through $L_j=(B_0^2/4\pi)c\pi r_L^2$.   The angular velocity of a nearly maximal Kerr black hole is $\Omega_H=c/2r_H\simeq c/r_s$. 
Assuming a rigid rotation with angular velocity $\Omega=\Omega_H/2$ for the Poynting dominated jet, we estimate $r_L\simeq c/\Omega=2r_s$.
Detailed analysis (Globus \& Levinson 2014) indicates $r_L=2.25 r_s$ for the outer light surface of 
an equatorial, force-free outflow from a Kerr black hole with $a=0.9$.    In what follows  we adopt this value.

{ From Equation (\ref{asymptotic_jet}) it is seen that substantial collimation is expected if $3\pi cp_{ws}z^2/2L_j\simgt1$.  
For a flat pressure profile, $p_{ws}\propto z^{-\kappa}$ with $\kappa<2$, this condition is always satisfied ultimately.   
For a steeper profile it can only be satisfied in a certain range, provided $p_{ws}$ is large enough (Lyubarsky 2009).   
For an adiabatic disk wind $p_{ws}\propto z^{-10/3}$ asymptotically, and so collimation can occur only in the region where
the streamlines diverge slower than radial, provided that at the collision point of  the jet and the innermost  streamline of the wind, $z=z_c$,  
the quantity $\beta_c\equiv 3\pi c z_c^2p_{ws}(z_c)/2L_j$ is not much smaller than unity.    A rough estimate of $\beta_c$
can be obtained as follows:   From Equation (\ref{p_ws}) we get for a supersonic wind $p_{ws}\simeq\rho_wu_w^2\sin^2\delta_w$. 
By employing Equations (\ref{geometry}), (\ref{massflux}), (\ref{pup1}) and (\ref{mass_cons}) below, and noting that $r_{jc}=r_{wc}\simeq r_0$, where $r_{jc}=z_c\tan\theta_{j,max}$ is the jet radius at $z_c$,  we obtain
\begin{equation}
\beta_c\equiv \frac{3\pi c\, z_c^2p_{ws}(z_c)}{2L_j}\simeq \frac{3\pi}{2\chi}\frac{sin^2\delta_w}{\sqrt{b_w}\ln(r_{out}/r_{in})}\frac{(z_c/r_s)^{3/2}}{\tan\theta_{j,max}}
\label{coll-crit}
\end{equation}
in terms of the jet-to-wind power ratio $\chi=L_j/L_w$.    In the examples presented below $\tan\theta_{j,max}=0.577$, $(z_c/r_s)^{3/2}\simeq50$,  and 
the value of $\sqrt{b_w}\ln(r_{out}/r_{in})$ is between $3$ and $5$.  Thus, collimation is expected when $\chi< 50\sin^2\delta_w$.  

Equation (\ref{asymptotic_jet}) is applicable well above the light cylinder, after a transition from 
free expansion to confined jet occurs.  In all the examples presented below our integration starts 
sufficiently above the light cylinder, at $r=$ a few $r_L$, where the asymptotic expansion of the transfield equation 
holds.  However, we apply Equation (\ref{asymptotic_jet}) in the entire regime, including the transition zone where the 
use of this equation may be questionable.   
Quite generally, at the transition zone the streamlines of the unconfinered jet are deflected either across a superfast  tangential shock, if the transverse 
expansion of the unconfined jet is super-magnetosonic, 
or smoothly through the formation of a compression wave if it is sub-magnetosonic (Levinson \& Begelman 2013).  In any case,
once the jet pressure near the contact adjusts to the external pressure Equation (\ref{asymptotic_jet}) applies. 
In situations where  the shocked wind pressure is comparable to the ram pressure of the magnetic jet, $\beta_c\sim1$, 
this adjustment is fast.  And at any rate, we find that when the shocked wind pressure is weak, viz.,  $\beta_c<<1$, the solution 
of Equation (\ref{asymptotic_jet})  maintains the shape of the freely expanding jet.      This is because the pressure profile in the 
shocked layer is never much shallower than $z^{-2}$,  as seen in the right panel of Figure \ref{f4}. 
We therefore conclude that  the accuracy of the results presented below is good . }
 
\subsection{\label{outer_wind}The outer disk wind }

In our simple treatment the disk wind is ejected from a region in the equatorial plane enclosed  
between two circles of radii $r_{in}$ and $r_{out}$ (see Figure \ref{f1}).
We assume  a shifted split-monopole geometry for the wind streamlines, with the origin located at  $(z,r) = (-d_1,d_2)$.
The inclination angle $\theta_w$ of a streamline emanating from a radius $r_0$ in the equatorial plane ($z=0$) 
satisfies $\tan\theta_w=(r_0-d_2)/d_1$.
This streamline intersects the shock at some height $z_w(r_0)$ defined implicitly through the relation
\begin{equation}
r_0 = \frac{r_w(z_w)-d_2}{1+z_w/d_1}+d_2\,,\label{r0}
\end{equation}
where $r_w(z)$ is the shock profile.   The cross sectional area of a flow tube enclosed between the 
angles $\theta_w$ and $\theta_w+d\theta_w$ at height $z$ is $dA(z)=2\pi(r_0+z\tan\theta_w)(d_1+z)d\theta_w$. 
The geometrical factor that gives the expansion of the wind streamlines is the ratio 
\begin{equation}
\frac{d A_w}{d A_0}=(1+z_w/d_1)\frac{r_w}{r_0}\,,
\label{geometry}
\end{equation}
where $dA_0=dA(z=0)$ and $dA_w=dA(z_w)$.   This ratio can be readily generalized to a Schwarzschild spacetime (see Levinson 2006).

For simplicity we consider a self-similar wind model for which the mass flux satisfies
 \begin{equation}
\dot{m}_w(r_0)=(\dot{m}_{\rm out}-\dot{m}_{isco}) \left(\frac{r_0}{r_{\rm out}}\right)^p\,\, \mbox{with} \,\, 0.5<p<1.5\,,
\label{massflux}
\end{equation}
where $\dot{m}_{isco}$ is the mass absorbed by the black hole, $\dot{m}_{\rm out}\equiv \dot{m}(r_{\rm out})$ the accretion rate at the outer 
boundary of the wind injection zone, and  $r_0$ denotes the cylindrical radius at the equatorial plane ($z=0$).  
A particular example is the two zone inflow-outflow solution model (ADIOS) 
proposed by Begelman (2012; see also Blandford  \& Begelman 1999, 2004)
in which most of the accreted plasma is lost through winds.   In general, the Bernoulli function {\it per unit mass} 
along a streamline emanating from a radius $r_0$ is given by 
\begin{equation}
B_w(r_0)=\frac{u_{w}^2(r_0)}{2}+\frac{{\cal L}_w^2}{2r_0^2}-\frac{{G}M_{BH}}{r_0}+\frac{\hat{\gamma}}{\hat{\gamma}-1}a_s^2(r_0)\label{Bernoulli_cons},
\end{equation}
here ${\cal L}_w$ is the specific angular momentum, to be neglected henceforth, $a_{s}=\sqrt{(p_w/\rho_w)}$ the 
isothermal sound speed, and $\hat{\gamma}$ the adiabatic index, chosen to be $5/3$ below.
For the self-similar model $r_0B_w(r_0)$ is independent of $r_0$.    In terms of the dimensionless  quantity $b_w=r_0 B_w/GM_{BH}$,
and the fiducial power 
\begin{equation}
\dot{E}_{p}\equiv{G}M_{BH}(\dot{m}_{\rm out}-\dot{m}_{isco})/r_{\rm out},
\label{dotE}
\end{equation}
the total wind luminosity can be expressed as
\begin{equation}
L_w= \left\{ \begin{array}{r@{\quad:\quad}l} \displaystyle {\dot{E}_{p}}b_w\ln\left(\frac{r_{\rm out}}{r_{\rm in}}\right) & 
\mbox{if} \,\, p = 1 \\ \displaystyle{\dot{E}_{p}}b_w\frac{p}{p-1}\left[1-\left(\frac{r_{\rm out}}{r_{\rm in}}\right)^{1-p}\right] & \mbox{if} \,\, p \neq 1\end{array} \right.
\label{pup1}
\end{equation}

The wind parameters just upstream of the shock cab be obtained in terms of the injected values at $z=0$ as follows: 
First, mass conservation along a given flux tube gives the mass flux at the shock,
\begin{equation}
\rho_w(z_w)u_w(z_w)=\rho_{w0}u_{w0}\frac{d A_0}{d A_w}\label{mass_cons},
\end{equation}
where $\rho_{w0}u_{w0}\equiv \rho_wu_w (z=0)$ is related to the differential mass loss rate through 
$\rho_{w0}u_{w0}={{\rm d} \dot{m}_w}/{(2\pi r_0\,{\rm d} r_0)}=p(\dot{m}_{\rm out}-\dot{m}_{isco})r_0^{p-2}/(2\pi\, r_{out}^p )$.  
Second, assuming adiabatic wind we have 
\begin{equation}
p_w(z_w)=p_{w0}[\rho_w(z_w)/\rho_{w0}]^{5/3}.\label{press_w}
\end{equation}
Finally, noting that the Bernoulli function is conserved along streamlines, $B_w(r_w)=B_w(r_0)$, and using
Equation (\ref{press_w}), we obtain $u_w(z_w)$.

\begin{figure*}
\includegraphics[width=16.5cm]{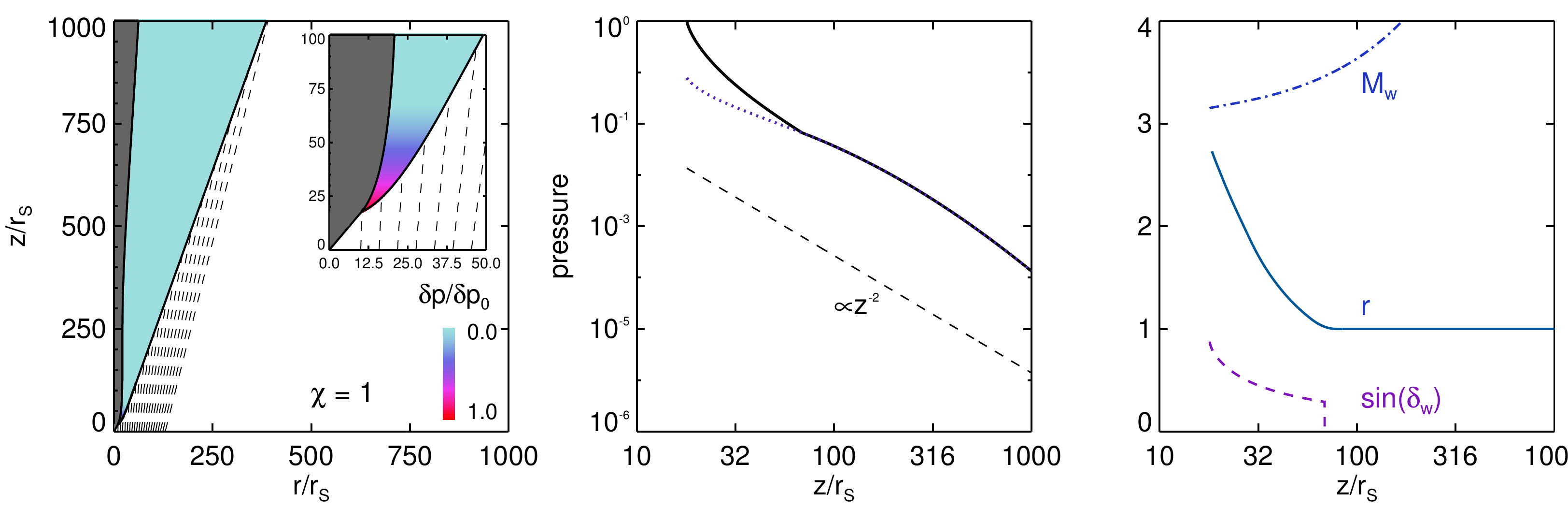}
\includegraphics[width=16.5cm]{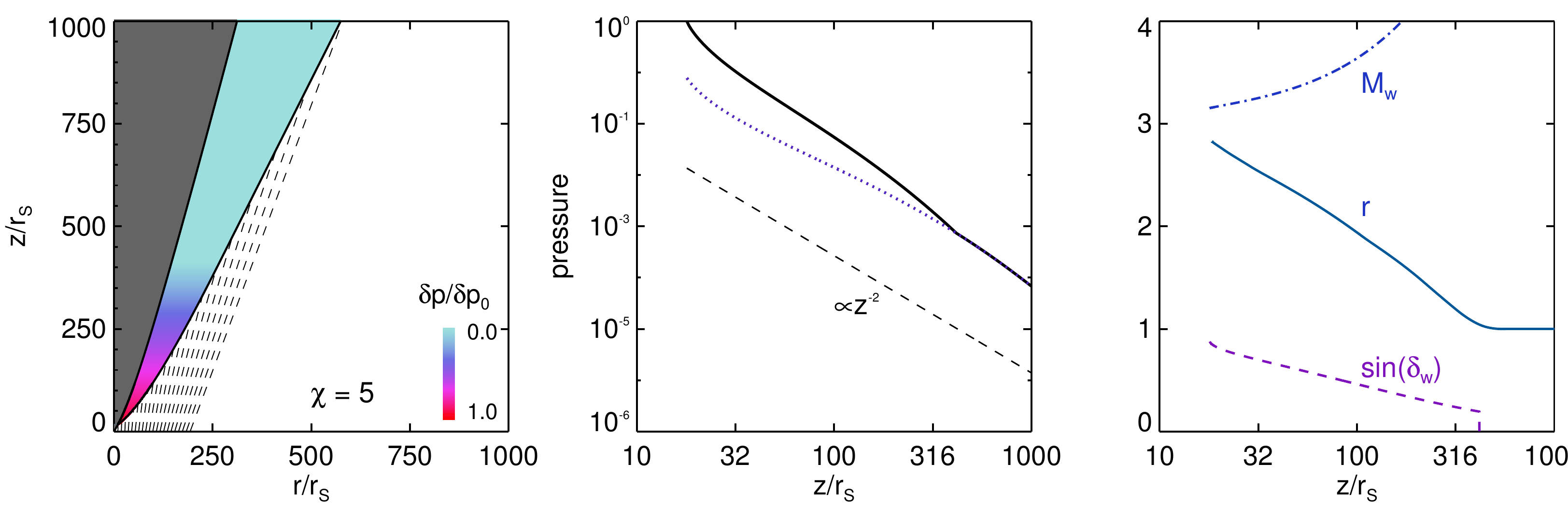} 
\caption{\label{f2} Multiflow structure produced by collision of a disk wind with a Poynting-flux dominated jet, for
two values of the power ratio $\chi$, as indicated .   The grey and coloured areas in the left panels mark the  
polar jet and the shocked wind layer, respectively. The colour code in the shocked wind layer
refers to the difference in kinetic pressures of the shocked and unshocked wind.
The dashed lines trace the streamlines of the unshocked wind.
Middle panel: profiles of the shocked (solid line) and unshocked (dotted line) wind pressure. 
The dashed line corresponds to  $p\propto z^{-2}$ and is shown for guidance. 
Right panel: Mach number of the disk wind just upstream of the shock (dotted-dashed line), shock compression ratio (solid line)
and sine of impact angle (dashed line).} 
\end{figure*}

\subsection{Method of solution}

The analysis can be considerably simplified upon assuming that the shocked wind flow is uniform,
in the sense that the corresponding fluid quantities, $p_{ws}$, $\rho_{ws}$ and the velocity $\vec{u}_{ws}$, 
are functions of $z$ only.  The boundary condition (\ref{BC_j}) then readily implies that the shocked
fluid velocity just downstream of the shock is parallel to the jet boundary.  The deflection angle in 
that case satisfies $\zeta(z)=\alpha_j(z)-\theta_w(z)$, where $\tan\theta_w=(r_w-d_2)/(z+d_1)$. 
With the above assumptions, the system is characterized by the following independent variables: $r_w(z)$, $r_j(z)$, 
$\alpha_j(z)$, $\delta_w(z)$, $p_{ws}(z)$, $\rho_{ws}(z)$ and $u_{ws}(z)$.  Their evolution is governed by 
Equations (\ref{BC_j}), (\ref{comp_r}),  (\ref{p_ws}),  (\ref{u_ws}), (\ref{dr_w/dz}), (\ref{deflect}) and (\ref{asymptotic_jet}), with the implicit input functions $\theta_w(r_w,z)$, $\rho_w(r_w,z)$, $p_w(r_w,z)$ and $u_w(r_w,z)$ derived in the preceding section.

We start the integration at the point where the collision between the jet and the innermost wind streamline ($\theta_w=\theta_{w,min}$) occurs. 
The coordinates of this point are given by
\begin{eqnarray}
z_c&=&\frac{\left(d_1\tan\theta_{w,min}+d_2\right)}{\left(\tan\theta_{j,max} - \tan\theta_{w,min}\right)}\,,\\
r_{jc}&=&z_c\tan\theta_{j,max}\,.\label{r_jc}
\end{eqnarray}
The slope of the jet surface, the deflection angle and the radius of the shock at the point of first impact are taken to be 
\begin{eqnarray}
\left(\frac{dr_j}{dz}\right)_{z_c}&=&\tan\theta_{j,max}\, ,\\
\zeta(z_c)&=&\theta_{j,max}-\theta_{w,min}\,,\\
r_{w}(z_c)&=&r_{jc}.
\end{eqnarray}
The initial impact angle $\delta_w(z_c)$ is then obtained upon choosing the weak shock solution to Equation (\ref{deflect}), and
the initial slope of the shock front upon substituting $\delta_w(z_c)$ and $\theta_w=\theta_{w,min}$ into Equation (\ref{dr_w/dz}).  The initial pressure, density and velocity are obtained likewise from Equations (\ref{comp_r}), (\ref{p_ws}) and (\ref{u_ws}), respectively. 

As explained above, our analysis is restricted to supersonic winds (albeit  mildly supersonic).  Thus,  
the collision of the wind and the magnetized jet always leads to formation of an oblique shock at
sufficiently small radii.  However, as the radius increases the shock weakens and eventually disappears.   
At  the transition from supersonic to subsonic collision the streamlines of the shocked wind layer smoothly joins the streamlines of 
the outer wind.    We find that essentially in all cases the shock surface becomes radial as the shock weakens.   
To simplify the analysis we assume that the shocked wind flow remains conical after the shock dies away, 
and that it is maintained in pressure balance with the external wind.   

\begin{figure*}
\begin{tabular}{cc}
   \includegraphics[width=8cm]{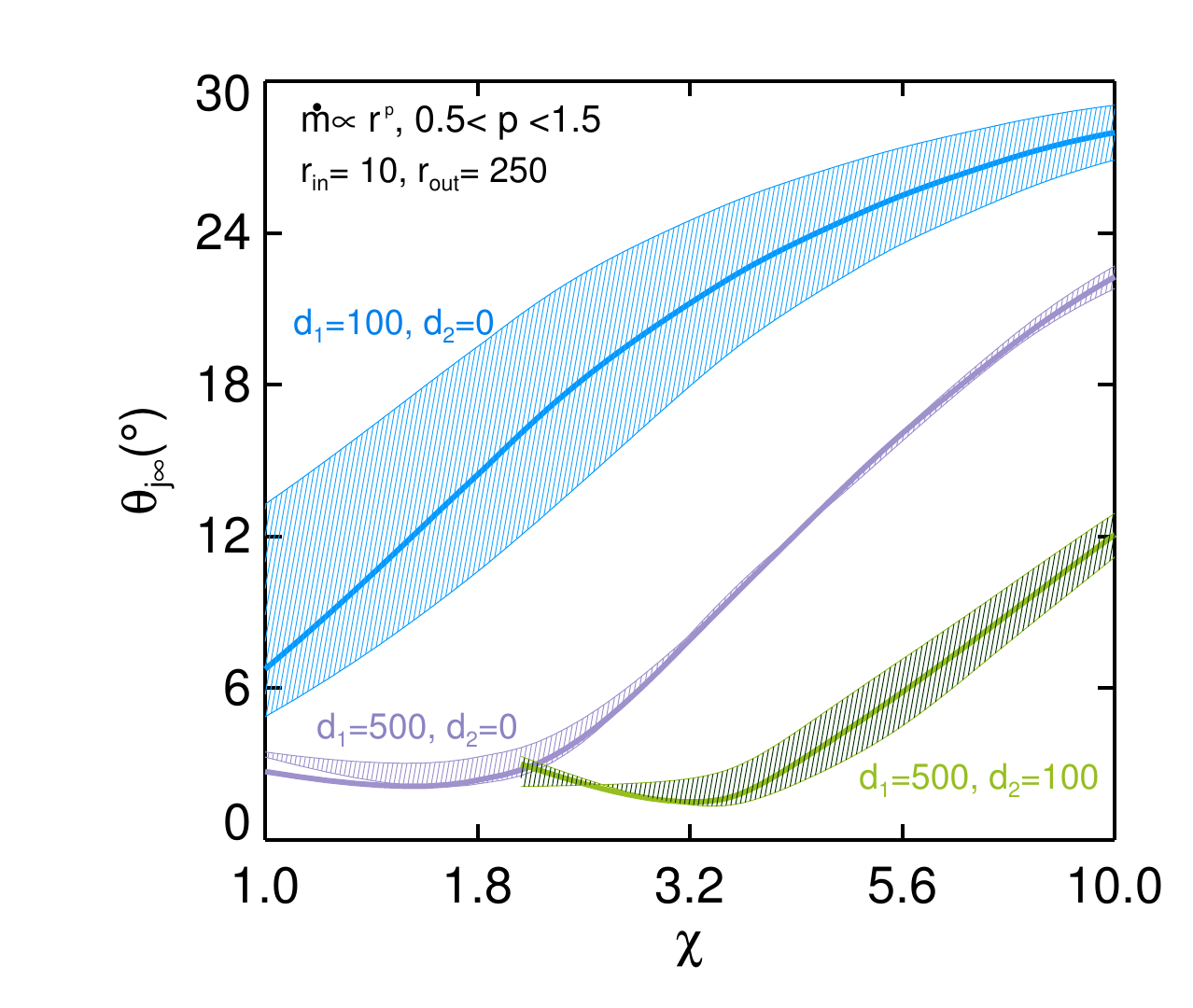} &
   \includegraphics[width=8cm]{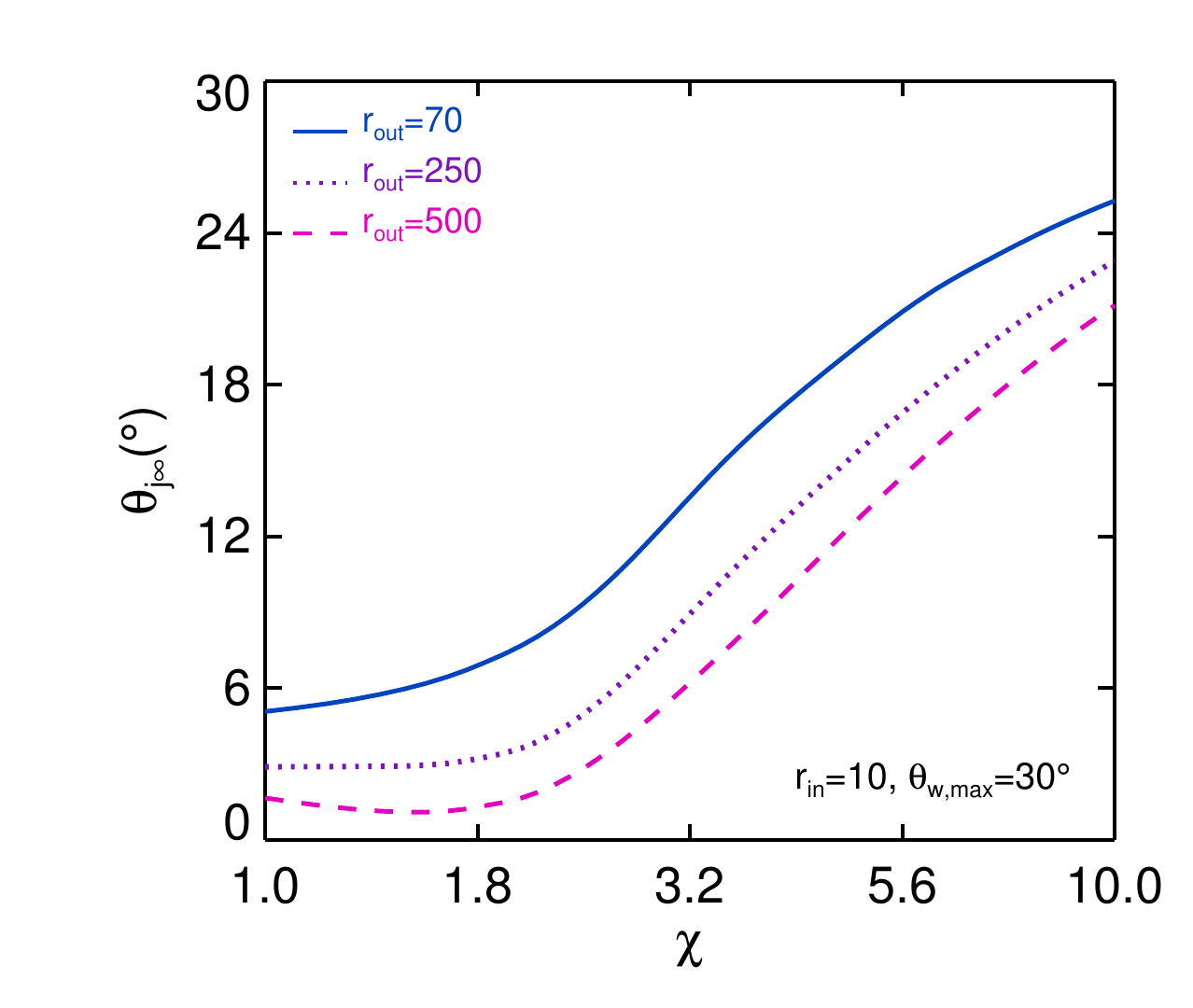} \\
\end{tabular}
 \caption{\label{f3}  A plot of the asymptotic opening angle $\theta_{j\infty}$ as a function of $\chi$ for different parameters of the  disk wind 
 model.  The coloured stripes around each curve in the left panel
correspond to the range of values $0.5\le p \le 1.5$ of the index of mass effluxion rate, with $p=0.5$ at the lower edge and $p=1.5$ at the upper edge. } 
\end{figure*}

\section{Results}

\begin{figure*}
\centering
\includegraphics[width=15cm]{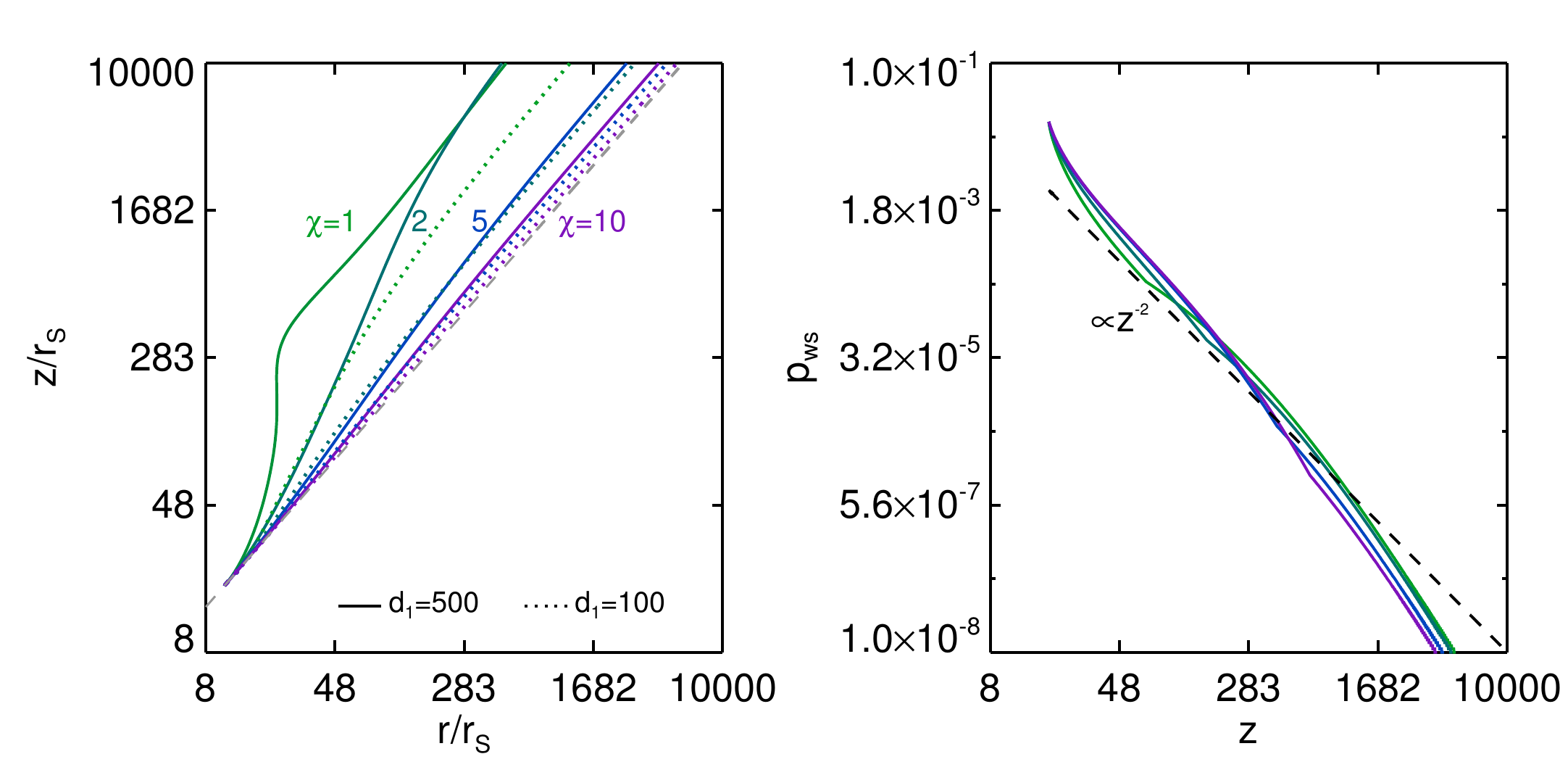}
\caption{\label{f4} Sample profiles of the jet radius $r_j$ (left panel) and shocked wind pressure $p_{ws}$ (right panel)
of some of the solutions computed in Figure \ref{f3}.   The solid lines correspond 
to the choice $d_1=500$, $d_2=0$ and different values of $\chi$, as indicated, and the dashed lines to $d_1=100$, $d_2=0$ and
same values of $\chi$.} 
\end{figure*}

In general, the profile of the inner jet depends on four main factors: the normalized wind power $\tilde{L}_w=L_w/\dot{E}_p$, 
the jet-to-wind power ratio, $\chi=L_j/L_w$,
the horizontal extent of the disk wind, $r_{out}$, and the inclination angles of the wind streamlines, which
for the geometry invoked here are determined by $d_1$ and $d_2$.  Results obtained for a broad range of these 
parameters are exhibited in figures \ref{f2}-\ref{f4}.

Two typical examples are displayed in Figure \ref{f2}, for $\chi=1$ (upper panels) and $\chi=5$ (lower pannels).
The initial opening angle of the inner jet in both examples is $\theta_{jmax}=30\degree$. The wind injection zone 
extends in the equatorial plane from $r_{in}=10 r_s$ to  $r_{out}=200 r_s$, with minimum and maximum inclination 
angles $\theta_{wmin}=0$ and  $\theta_{wmax}=22^\circ$ ($d_1=500$, $d_2=0$ in both cases), and with index of 
mass effluxion rate $p=1$ (see Equation (\ref{massflux})).
The disk wind in these examples is mildly supersonic, as seen in the right panels in Figure \ref{f2}.
The left panels display the evolution of the jet radius and the shocked wind.  The streamlines of the unshocked 
wind are delineated by the dashed lines. The colour code in the shocked layer refers to the difference between the 
pressures downstream and upstream of the shock normalized to its value at the collision point:
$\delta p/\delta p_c \equiv [p_{ws}(z) - p_{w}(z)]/[p_{ws}(z_c) - p_{w}(z_c)]$.  
As can be seen, at small distances the shock is relatively strong, and the  inner jet is essentially
confined by the ram pressure of the disk wind.
As $z$ increases the impact angle $\delta_w$ decreases and the shock weakens and eventually disappears, 
at which point the jet becomes confined by the kinetic pressure of the unshocked wind.   
This transition from supersonic to subsonic collision is seen more clearly in the middle panels that delineate the 
pressure distribution in the shocked (solid) and unshocked (dotted) disk wind, and in the right panels where the compression ratio $r$ is plotted.  
The oscillatory behaviour of the inner jet seen in the upper left panel is typical in cases where the wind power approaches
the jet power ($\chi\simlt2$).   The focusing of the jet results from the shallower pressure profile at the jet boundary ($p_{ws}\propto z^{-\kappa};\, \kappa<2$) in the region $50\, r_s<z<300\, r_s$.   
For larger values of $\chi$ the collimation is gradual.  In all cases the inner jet becomes ultimately conical
with a final half opening angle $\theta_{j\infty}$ that depends on $\chi$ and the geometry of the disk wind. 
For the examples exhibited in Figure \ref{f2} we obtained $\theta_{j\infty}\approx3\degree$ for $\chi=1$ and $\theta_{j\infty}\approx15\degree$ for $\chi=5$.

To explore how the geometry and power distribution of the disk wind affects the collimation of the magnetized jet, we derived solutions corresponding to different choices of the parameters $d_1$, $d_2$, the index of mass effluxion rate $p$, and the normalized wind power $\tilde{L}_w$.   
Figure \ref{f3} depicts the final half opening angle of the 
jet $\theta_{j\infty}$ as a function of $\chi$, for different choices of the parameters, as indicated.  The coloured stripes around each curve in the left panel
correspond to the range of values $0.5\le p \le 1.5$, with $p=0.5$ at the lower edge and $p=1.5$ at the upper edge.   The left panel exhibits 
results obtained for a fixed $r_{out}$ and different choices of $d_1$ and $d_2$.   
As seen, better collimation is achieved for larger values of $d_1$ (i.e. less inclined streamlines). 
The reason is that at distances $z<d_1$ the streamlines of the disk wind diverge slower than conical, roughly as $\Delta A\propto z$,
giving rise to a sufficiently flat profile of the confining pressure $p_{ws}(z)$ (slightly flatter than $z^{-2}$) to maintain good collimation up to $z\simeq d_1$.
Beyond that distance the streamlines diverge conically ($\Delta A\propto z^2$), leading to a steeper decline of $p_{ws}(z)$  (faster than $ z^{-2}$), at
which point the inner jet becomes radial.   This behaviour is consistent with the trends found in Lyubarsky (2009).   However,  it should be noted
that in the situations considered here the profile of the confining pressure is regulated by the collimation of the inner jet,  in difference to the case of a prescribed pressure profile considered in Lyubarsky. 
Increasing $d_2$ for a fixed $d_1$ also leads to a better collimation.  This is because the impact angle of the wind streamlines, $\delta_w$, and the 
consequent pressure in the shocked layer, $p_{ws}$, increase with increasing $d_2$.  The profiles of the jet radius $r_j$ and the 
confining pressure $p_{ws}$ corresponding to some of the cases computed in Figure \ref{f3} are
depicted in Figure \ref{f4}.  We use a logarithmic scale to show the transition from a  parabolic to a conical structure.    
  
Somewhat better collimation is achieved also in case of more extended winds (that is, larger $r_{out}$) with the same
inclination of streamlines, as can be seen in right panel of Figure \ref{f3}, where the final opening angle $\theta_{j\infty}$ is plotted
as a function of $\chi$ for different values of $r_{out}$ but the same $\theta_{w,max}$.
The reason is that the pressure of the magnetic jet drops much faster with $r$ than the ram pressure of the 
colliding wind, and so even though collimation starts at a larger $z$ it is maintained over relatively 
larger scales and, therefore, more effective.    In general,  for given wind power and geometry, the distance over 
which the jet collimates  increases with $r_{out}$.

{ For all cases explored above substantial collimation is observed for $\chi \simlt 10$.  This is consistent with the limit derived in 
Equation (\ref{coll-crit}).  Weaker winds can still collimate the jet
if the inner streamlines have larger inclination angles, as in the case $d_2=100$, $d_1=500$ shown in Figure \ref{f3}. 
This is because the pressure profile of the shocked wind is shallower over larger distances. }

\section{Implications for magnetic dissipation}

The conversion of magnetic energy to kinetic energy in Poynting dominated jets is yet an unresolved issue. 
A plausible dissipation mechanism commonly invoked
is magnetic reconnection (Romanova \& Lovelace 1992, Levinson \& van Putten 1997, 
Drenkhahn \& Spruit 2002; Lyutikov \& Blandford 2003; 
Giannios \& Spruit 2007; Lyubarsky 2010; McKinney \& Uzdensky 2012, Levinson \& Begelman 2013, 
Bromberg \& Tchekhovskoy 2016; Levinson \& Globus 2016).   This mechanism requires
formation of small scale magnetic domains with oppositely oriented magnetic field lines.   Such 
structures may inherently form during outflow injection (e.g., Drenkhahn \& Spruit 2002; Levinson \& Globus 2016)
, or result from current-driven instabilities induced during the propagation of the 
jet (Mignone et al. 2010, Mizuno et al. 2012, O'Neill et al. 2012, Guan et al. 2014, Bromberg \& Tchekhovskoy 2016).

Recently, Bromberg \& Tchekhovskoy (2016) observed a rapid growth of internal kink modes, followed by magnetic dissipation 
through small-angle reconnection, in 3D MHD simulations of jet propagation 
in the envelope of a GRB progenitor.   The growth of the instability occurs at the recollimation nozzle, where the jet, following 
a free expansion stage, is focused by the over pressured cocoon.  The rapid growth of kink modes observed in their simulations
is partly due to magnetic field compression by the ram pressure exerted on the jet's head
that enhances the ratio of toroidal to poloidal magnetic field strengths [see Equation (\ref{l_kink}) below].
Whether such compression is crucial in general for the instability to develop is unclear at present.  One naively 
expects that jet sections in which the Alfv\'en crossing time is much shorter than the expansion time of the flow will be unstable.
Recollimation nozzles should be particularly kink unstable. 

The e-folding length of internal kink modes is approximately (e.g., Appl et al. 2000)
\begin{eqnarray}
\lambda_{kink}\simeq \frac{2\pi r_j \Gamma_j}{\beta_A}\frac{B^\prime_p}{B^\prime_\phi},
\label{l_kink}
\end{eqnarray}
here $\Gamma_j$ is the Lorentz factor of the jet, $B_p^\prime$ and $B^\prime_\phi$ are the poloidal and toroidal components of the 
proper magnetic field, respectively, and $\beta_A\simeq1$ is the local Alfv\'en speed (in units of $c$).   Dissipation of the magnetic 
field occurs over a length scale of a few $\lambda_{kink}$.  Well above the light cylinder 
$B_p^\prime\simlt B_\phi^\prime$, so that growth of the instability is anticipated in regions 
where $z\simeq10\lambda_{kink}\simeq60 r_j\Gamma_j$.   
Based on this simple estimate, we speculate that the internal kink instability  might
develop in the recollimation nozzle of the jet exhibited in the upper left panel of Figure \ref{f1}.  
This situation is quite typical for confinement by disk winds when $\chi\simlt1$.  Whether this eventually leads to magnetic
dissipation can only be explored with full 3D MHD simulations.   Jets confined by weaker winds, as in the 
examples shown in the bottom left panel of Figure \ref{f2} and in Figure \ref{f5} are expected to be stable.

\section{Application to M87}\label{sec:M87}

Recently, special efforts have been made combining VLBA and EHT observations, 
to resolve the jet formation region in the elliptical galaxy M87
(see e.g., Ly et al. 2007, Hada et al. 2011, Asada \& Nakamura 2012, Doeleman et al. 2012). 
The observed parabolic structure ($r\propto z^{0.58}$) maintains over several decades of radius, from $z\sim50 r_s$ up to 
$z\sim10^5 \,r_s$, above which it becomes conical.   In the inner region, $z<50 \,r_s$, the radio jet  seems to show weak collimation,
roughly $r\propto z^{0.8}$, although this fit is uncertain due to the scarce data.   In addition, velocity measurements seem to indicate a 
multi-flow structure in the collimation zone,
with some radio components moving at sub-relativistic speed (Kovalev et al. 2007) and a few at mildly relativistic speed (see, e.g., Asada
et al. 2014 for a compilation of measurements).   Measurements of apparent velocities on larger scales ($z>10^4 \,r_s$)  
indicate relativistic motion. The multi-flow structure observed on small scales is expected in our model; the slow components observed
with the VLBA by Kovalev et al. (2007) are attributed to emission by the sub-relativistic shocked flow, whereas the fast components to emission 
from the inner Poynting dominated jet.   Variations of this idea have been considered earlier  by McKinney (2006), 
Gracia et al. (2009), Nakamura \& Asada (2013), and Moscibrodzka et al. (2016).   

\begin{figure*}
\centering
\includegraphics[width=15cm]{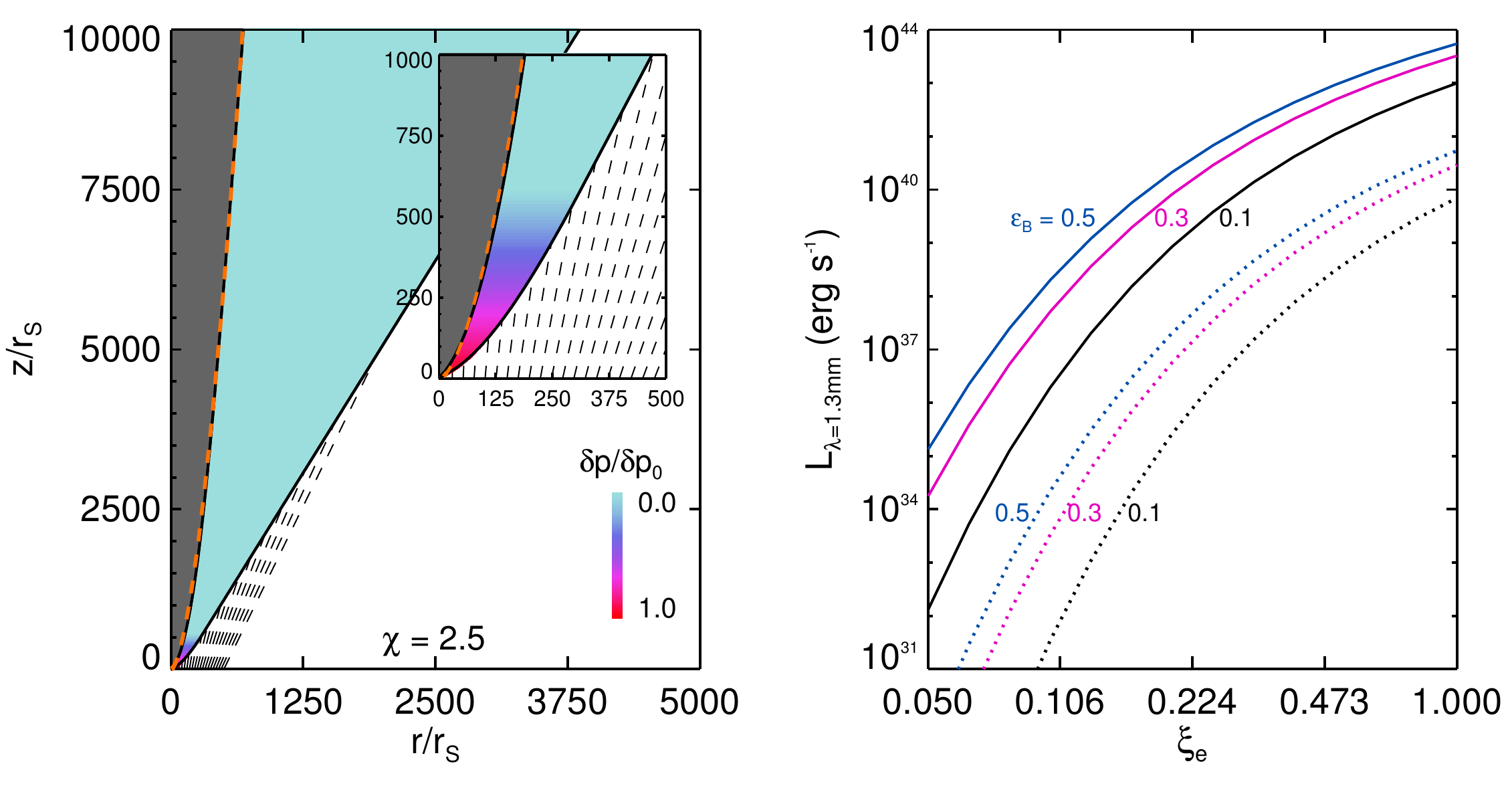}
\caption{\label{f5} Left panel: Collimation profile  of the M87 jet produced by the colliding outflows 
model with $\chi=2.5$ and $r_{out}=500\, r_s$. 
The red dashed line corresponds to the profile of the observed radio jet, as reported in Asada \& Nakamura (2012).  
Right panel:  Synchrotron luminosity at 1.3 mm emitted from the shocked layer, 
for different values of $\xi_e$, $\epsilon_B$ and $L_j$.  The luminosity is defined as $L_{\lambda=1.3mm}=\nu L_\nu(\nu=230 \rm GHz)$, where 
$L_\nu$ is the spectral luminosity.  The solid and dotted lines correspond to a jet power of $L_j=10^{44}$ erg/s and 
$L_j=10^{43}$ erg/s, respectively. } 
\end{figure*}

In order to reproduce the entire collimation profile by a disk wind, the wind injection zone must extend to relatively large radii. 
An example is shown in the left panel of Figure \ref{f5}, where the parabolic profile observed over 
nearly 4 decades in radius (red dashed line) is reproduce
using $\chi=2.5$, $r_{out}=500\, r_s$, $\theta_{w,max}=25 ^\circ$, $p=1$.    Whether such an extended wind is realistic is unclear
(note however, that the wind power is dominated by the inner regions.  For our choice of $p=1$  about $70$ percents of the 
power emerges from within $r=100 \,r_s$).    If not, a different pressure source with an appropriate profile should be present in the collimation zone. 
In the latter case, a strong wind from the inner disk regions should be avoided, as otherwise 
it will cause strong collimation already at small radii, within several hundreds $r_s$.   We note that the parabolic profile in this specific example 
extends down to the base of the jet.  This is inconsistent with the weaker collimation claimed to be seen below about $50 \,r_s$.
{ Such a transition can, in principle, be reproduced in our model by reducing the power emerging along the inner streamlines, such that beneath 
the transition radius the ram pressure of the disk wind will be smaller than that of the jet.   
However, it is more likely that the radio image observed on those scales does not reflect the jet profile.   As explained below, the radio emission 
probably originates from the jet sheath, the profile of which
differs from that of the jet boundary, as can be seen in the inset in the left panel.   At sufficiently small radii the emission profile
conceivably traces the outer sheath boundary.  In fact, the shock profile we obtained within this region is 
quite consistent with the observations.   Further up, the radio emission is expected to originate from 
a thin, hot layer within the sheath adjacent to the jet boundary (see discussion below), thereby reflecting the jet profile.    
Thus, the observed transition may not be associated with a change in the jet profile, but rather involves 
the distribution of the electron temperature and magnetic field in the sheath. }

The origin of the radio emission observed on EHT scales is unclear at present.    If the source of this emission is the inner, Poynting dominated  jet 
it would require magnetic field dissipation just above the light cylinder.   This seems unlikely.  The observed collimation profile
indicates that the jet is kink stable (see discussion in preceding section).    It could be that an unstable magnetic field configuration is 
established  during the injection 
of the jet, e.g., owing to advection of asymmetric magnetic fields by the accretion flow into the ergosphere.   However, the dissipation scale
of such a quasi-striped configuration is $r_{diss}\simeq 10\,r_s\,\Gamma_j^2>>r_s$ (e.g., Drenkhahn \& Spruit 2002; Levinson \& Globus 2016).
An alternative mechanism  is emission by electrons accelerating in a spark gap located around the stagnation  surface of the plasma 
double-flow (Levinson 2000).  
While this process can successfully account  for the variable TeV emission (Levinson 2000 ; Neronov \& Aharonian 2007; Levinson \& Rieger 2011;
Broderick \& Tchekhovskoy 2015; Hirotani \& Pu 2016), it 
would require the plasma density in the gap to exceed the Goldreich-Julian value by several orders of magnitudes in 
order to produce the observed radio emission (Broderick \& Tchekhovskoy 2015). 
Such high densities are implausible, at least in static gaps, 
as screening of the gap should occur when the charge density approaches the Goldreich-Julian value (see additional
arguments in Hirotani \& Pu 2016).    Since plasma in a starved magnetosphere of a black hole (unlike in a pulsar) can only be supplied through
pair production, the spark gap is most likely intermittent (Levinson et al. 2005; Timokhin \& Arons 2013). 

As hinted above, the core emission can be produced in the hot sheath flow surrounding the inner jet.   Under the conditions anticipated
in a radiatively inefficient accretion flow (RIAF), the dominant radiation mechanism at millimetre wavelengths is synchrotron emission by 
relativistic thermal electrons.   
The electron temperature in the RIAF depends on the electron-ion coupling, and is highly uncertain. 
For a viable range of  microphysical conditions it can be relativistic, $kT_e>>m_ec^2$, in the inner regions.    
In order that the sheath emission will dominate
the emergent spectrum, the electron temperature in the sheath flow must be considerably higher than the electron 
temperature in the accretion disk (Yuan et al. 2002;  Moscibrodzka \& Falke 2013).   This requires additional 
heating of electrons in the sheath flow, e.g., due to a stronger coupling with ions.   
The microphysical processes that might give rise to such heating are not identified at present, so that $T_e$ cannot be computed
from first principles.   It is commonly used as  a free parameter in emission models, and we will adopt the same approach. 

We do not attempt to compute detailed spectra and images, but merely to check whether the conditions required to reproduce the collimation profile 
can be made consistent with radio observations.  The reason is that the shape of the spectrum and the 
resultant image depend sensitively on the temperature and magnetic field distribution in the shocked layer, 
that cannot be computed accurately in our model.    
To be concrete,  our analysis assumes that fluid quantities in the shocked layer depend only on $z$.    
While this is a reasonable assumption for the total pressure, the temperature and magnetic field 
might vary significantly across the sheath.    The point is that fluid elements moving along streamlines which are closer 
to the jet boundary originated from smaller radii and, therefore, have a higher temperature than that of fluid elements 
that joined the sheath at larger radii.     In addition, the ratio $T_e/T_p$ itself may vary across the sheath.  
For instance, with the prescription for $T_e/T_p$ adopted in Moscibrodzka et al. (2016) it is highest near the jet boundary.
Consequently, we anticipate that fluid elements moving along streamlines adjacent to the jet boundary will dominate the emission.    

For illustration, we present in the right panel of Figure \ref{f5} the synchrotron luminosity at $\lambda=1.3$ mm 
emitted by thermal electrons in the shocked layer, for different values of the electron-to-proton 
temperature ratio, $\xi_e=T_e/T_p$, magnetic field-to-gas pressure ratio, $\epsilon_B=p_{mag}/p_{ws}$, 
and fiducial jet power of $L_j=10^{44}$ erg/s (solid lines) and $L_j=10^{43}$ erg/s  (dashed lines).
The luminosities were computed using the solution presented in the left panel. 
The electron temperature in the jet sheath is assumed to be a fixed fraction of the proton temperature, $T_e=\xi_e T_p$, with $\xi_e$ constant. 
The proton temperature is computed from the dynamical model using $kT_p=m_pp_{ws}/\rho_{ws}$ (see further details in appendix \ref{appB}).    
We find that the major contribution to the synchrotron luminosity comes from the region $z\simlt 15 r_s$.   This is an
artefact of our assumption that the electron temperature is uniform across the sheath (see discussion above).    A more complete treatment of the temperature 
distribution would lead to a shallower longitudinal emission profile (from a thinner layer near the jet boundary). 
As seen, larger values of $\xi_e$ and $\epsilon_B$ are needed to produce the observed luminosity, $L_{\lambda=1.3mm}\simeq10^{41}$ erg/s, 
for a smaller jet power.    Increasing, instead, the wind power would lead to a faster collimation, inconsistent with the observed profile. 
We note that our assumption that the disk wind  is supersonic (${\rm M}_w\sim 3$ in the example shown in Figure \ref{f5}) implies that the proton temperature
in our model is underestimated by a factor of a few.   For a subsonic wind with the same power and 
geometry, the synchrotron luminosity produced for a given choice of $\xi_e$, $\epsilon_B$ and $L_j$ 
should be considerably higher than the values exhibited in Figure \ref{f5}.   Yet, we do not expect the collimation 
profile to be altered significantly, as it merely depends on the pressure profile of the disk wind.

\section{Summary}

We computed the multi-flow structure produced through the collision between a disk wind and a Poynting-flux dominated jet 
emanating from the ergosphere of a rotating black hole, using a semi-analytic approach, and treating the injection of the disk 
wind as a boundary condition in the equatorial plane.    We obtained solutions for a range of wind parameters, including total 
power, power distribution, the inclination angle of  streamlines and the extent of the wind injection zone.
Quite generally we find that jet collimation occurs when the ratio between the wind power $L_w$ and the jet power $L_j$ exceeds 0.1 roughly.   
The collimation occurs on scales at which the disk wind expands slower than radial.   In this region the pressure profile in the sheath 
flow surrounding the jet is approximately $p_{ws}\propto z^{-2}$ for a wide range of wind parameters.   At sufficiently large distances the 
pressure profile steepens and the jet becomes conical, with a finite opening angle that depends on the wind power and geometry. 
For mild winds, $0.1\,L_j\simlt L_w < 0.5\,L_j$, the collimation is gradual.   For  stronger winds, $L_w\sim L_j$,  
the collimation is fast followed by jet focusing.  The jet in those cases passes through a collimation  nozzle above which it
is maintained narrow over many light cylinder radii,
suggesting that it might be susceptible  to the current-driven kink instability.  We speculate 
that this may lead to magnetic field dissipation.    Otherwise, some other dissipation mechanism is 
required to convert magnetic energy
into kinetic energy.   One possibility is that the initial magnetic field configuration is unstable.   

As a particular case, we applied our model to the M87 jet and found that the profile of the radio jet observed over several decades in
radius can be reproduced provided that the confining wind is sufficiently extended, and that its power is roughly one third of the jet power. 
We argued that the radio emission observed on sub-parsec scales most likely originates from a thin, hot layer in the sheath flow adjacent to the jet boundary,
and speculated that the observed change in the collimation profile at roughly $50\, r_s$ can be attributed to the distribution of electron
temperature and magnetic field in the sheath.  Whether the wind from the RIAF satisfies the requirements needed to reproduce the observed 
collimation profile is unclear at present.    If not, other pressure source with a profile $\propto z^{-2}$ must be present in the jet collimation 
zone.   We note that in this case a substantial disk wind from the inner regions should be avoided, as it would lead to an inconsistent profile. 
For instance, a wind injected from radii $r_0<100 \,r_s$ with a power exceeding $0.3\,L_j$ would lead to a faster collimation than observed. \\

{ The observed collimation profile indicates that the M87 jet, if indeed magnetically dominated, is kink stable.  Magnetic 
dissipation would then require pre-existing asymmetries.   Horizon scale polarization maps may reveal the structure of
the magnetic filed at the base of the jet, and may enable estimation the fraction of the injected Poynting power available for dissipation.
While magnetic reconnection can account for the radio and X-ray  emission emitted by the jet, it is unclear whether it  can explain the 
rapid TeV variability observed on several occasions.   Misaligned minijets (Giannios et al. 2010), if indeed form in reconnection events,
may account for the variable TeV emission.   Alternatively, the TeV emission may originates from a spark 
gap (Levinson \& Rieger 2011; Hirotani \& Pu 2016), or result from jet-cloud interaction (Barkov et al. 2012).}\\

We thank Yuri Lyubarsky for enlightening discussions.  
NG acknowledges the support of the I-CORE Program of the Planning and Budgeting Committee and The Israel Science Foundation (grant 1829/12) 
and the Israel Space Agency (grant 3-10417). AL acknowledges the support of The Israel Science Foundation (grant 1277/13).

\onecolumn
\appendix
\section{ Jump conditions across unmagnetized oblique shocks}\label{appA}
Equation (\ref{deflect}) gives the  deflection angle $\zeta$  in terms of  the freestream Mach number
${\rm M}_w=v_w/a_s$, the impact angle $\delta_w$  
and the ratio of specific heats $\hat{\gamma}$. This relation may be written in a form of a cubic equation:
\begin{equation}
\left(1+\displaystyle\frac{\hat{\gamma}-1}{2}{\rm M}_w^2\right)\tan\zeta\tan^3\delta_w-({\rm M}_w^2-1)\tan^2 \delta_w +
\left(1+\displaystyle\frac{\hat{\gamma}+1}{2}\right)\tan\zeta \tan\delta_w + 1 = 0 .
\label{angle_eq}
\end{equation}
Equation (\ref{angle_eq})  admits three real roots, and an explicit expression for the oblique shock angle $\delta_w$ can thus be derived:
\begin{equation}
\delta_{w} =\tan^{-1}\left[\frac{{\rm M}_w^2-1+2f_1({\rm M}_w,\zeta)\cos\left(\displaystyle\frac{ k\pi+\cos^{-1}(f_2({\rm M}_w,\zeta))}{3}\right)}{3\left(1+\displaystyle\frac{\hat{\gamma}-1}{2}{\rm M}_w^2\right)\tan\zeta}\right]\, {\rm with }\,\, {k=0,2,4}, 
\end{equation}
with the functionals $f_1$ and $f_2$  given by
\begin{eqnarray}
f_1 ({\rm M}_w,\zeta)&=&\sqrt{\left({\rm M}_w^2-1\right)^2-3\left(1+\displaystyle\frac{\hat{\gamma}-1}{2}{\rm M}_w^2\right)\left(1+\displaystyle\frac{\hat{\gamma}+1}{2}{\rm M}_w^2\right)\tan^2\zeta}\,,\\
f_2 ({\rm M}_w,\zeta)&=&\frac{\left({\rm M}_w^2-1\right)^3-9\left(1+\displaystyle\frac{\hat{\gamma}-1}{2}{\rm M}_w^2\right)\left(1+\displaystyle\frac{\hat{\gamma}-1}{2}{\rm M}_w^2+\displaystyle\frac{\hat{\gamma}+1}{4}{\rm M}_w^4\right)\tan^2\zeta}{f_1^3 ({\rm M}_w,\zeta)}\,.
\end{eqnarray}

The two physically meaningful roots are the solutions $k=0$ and $k=4$, corresponding to a strong and a weak shock,
respectively (the root $k=2$ results in a decrease in entropy, and is unrealistic in steady state situations).
In this paper we use the weak shock solution $k=4$, that gives the appropriate asymptotic behaviour.

\section{Synchrotron luminosity from the shocked wind}\label{appB}
We assume that the electron temperature $T_e$ is a fraction $\xi_e$  of the proton temperature $T_p$:
\begin{equation}
\theta_e =\xi_e  \theta_p (m_p/m_e),
\end{equation}
where $\theta_e\equiv k_B T_e/(m_e c^2)$ and $\theta_p\equiv k_B T_p/(m_p c^2)$  are the normalized electron and proton temperatures.
In the optically thin limit, the spectrum of synchrotron emission by a relativistic Maxwellian electron distribution of total density $n_e$, moving 
in the direction $\theta$, is (Pacholczyk 1970)
\begin{equation}
j_\nu d\nu = 4.43\,10^{-30}\frac{4\pi n_e \nu}{K_2(1/\theta_e)}I\left(\frac{x_M}{\sin\theta}\right)d\nu\,[\textrm{erg. cm}^{-3}\textrm{s}^{-1} \textrm{Hz}^{-1} ],
\label{emiss}
\end{equation}
where
\begin{equation}
x_M\equiv\frac{2\nu}{3\nu_L\theta_e^2},\, \, \nu_L\equiv\frac{eB}{2\pi m_e c}.
\end{equation}
Here $\nu_L=2.8\cdot10^6\,B$ (Hz), where $B$ is the magnetic field in Gauss,
and $I(x_M)$ can be estimated by the following fitting function (Mahadevan et al. 1995),
\begin{equation}
I(x_M)=2.5651\left(1+\frac{1.92}{x_M^{1/3}}+\frac{0.9977}{x_M^{2/3}}\right)\exp(-1.8899\,x_M^{1/3}).
\end{equation}
The emissivity of an isotropic electron distribution is the integral of (\ref{emiss}) over $\theta$.   
It is given by Equation (\ref{emiss}) with  $I(x_M/\sin\theta)$ replaced by $I'(x_M)$, where the new fitting function is (Mahadevan et al. 1996)
\begin{equation}
I'(x_M)=\frac{4.0505}{x_M^{1/6}}\left(1+\frac{0.40}{x_M^{1/4}}+\frac{0.5316}{x_M^{1/2}}\right)\exp(-1.8899\,x_M^{1/3}),
\end{equation}
with a maximum error of 0.015 for $I'(x_M)$.

The magnetic pressure in the shocked layer is assumed to be a constant fraction $\epsilon_B$ of the gas pressure, yielding 
$B=({8\pi\epsilon_b p_{ws}})^{1/2}$.   
From Equation (\ref{emiss}), it follows that the synchrotron spectral luminosity integrated over the entire shocked layer is
\begin{equation}
\int_{\Lambda} dL_\nu(z)=\int_{z_{min}}^{\infty} j_\nu(z) \pi (r_w^2-r_j^2) dz,
\end{equation}
where $\Lambda$ is the total volume of the shocked layer.  The luminosity at some frequency  $\nu$ is defined as $L_{syn}(\nu)=\nu L_\nu$.  
The electron number density in the shocked layer is given by $n_e(z)=\rho_{ws}/m_p$, where $\rho_{ws}$ is  computed from the dynamical model. 
The ratio of the synchrotron and the jet luminosities can then be expressed as
\begin{equation}
\frac{ L_{syn}(\nu)}{L_j}=4.43\,10^{-30} \frac{r_s}{m_e c^3}\frac{4\pi \nu^2}{\chi\ln\left({r_{out}}/{r_{in}}\right)} \int_{z_{min}}^{z_{max}} \frac{ \bar{p}_{ws}(z) I'\left[{x_M(z)}\right]}{\theta_e(z)\,K_2(1/\theta_e(z))}(r_w^2-r_j^2) dz,
\end{equation}
in terms of the normalized pressure $\bar{p}_{ws}(z)=\pi r_s^2\, c\, p_{ws}(z)/\dot{E}_p$, where $\dot{E}_p$ is defined in Equation (\ref{dotE}).  All 
lengths  inside the integral are measured in units of $r_s$.

\end{document}